\documentclass[prb,twocolumn,showpacs,superscriptaddress]{revtex4}
\usepackage{mathptmx}
\usepackage[T1]{fontenc}
\usepackage[latin9]{inputenc}
\usepackage{verbatim}
\usepackage{amsmath}
\usepackage{graphicx}
\usepackage{epsfig,wrapfig}
\usepackage[usenames,dvipsnames]{color}

\newcommand{\bq}{\begin{mathletters}}
\newcommand{\eq}{\end{mathletters}}
\newcommand{\beq}{\begin{eqnarray}}
\newcommand{\eeq}{\end{eqnarray}}
\newcommand{\beqq}{\begin{eqnarray*}}
\newcommand{\eeqq}{\end{eqnarray*}}

\newcommand{\ee}{{\bf E}}

\begin{document}

\title{Three-dimensional coherence matrix and degree of polarization}

\author{Andrey V. Novitsky}
\email{andrey.novitsky@tut.by} \affiliation{Department of
Theoretical Physics and Astrophysics, Belarusian State University,
Nezavisimosti Avenue 4, 220030 Minsk, Belarus}

%\author{Leonid M. Barkovsky}
%\affiliation{Department of Theoretical Physics and Astrophysics,
%Belarusian State University, Nezavisimosti Avenue 4, 220030 Minsk,
%Belarus}

\begin{abstract}
Inspecting three-dimensional partially polarized light fields we
show that there is no unambiguous correspondence between the
three-dimensional field and coherence matrix (or light beam tensor).
Therefore, it is needed to clarify the definition of unpolarized
light. We believe that unpolarized field should be treated as light
of equiprobable polarizations similar to the case of two-dimensional
light. Then degree of polarization bridges two definitions of the
three-dimensional degrees of polarization known in literature. We
reveal that only 6 Stokes parameters are sufficient to describe the
coherence matrix. All these parameters can be retrieved from the
in-plane measurements of two-dimensional coherence matrices.
\end{abstract}


%%% ======================================================================

\maketitle

\section{Introduction}

The well-established concept of polarization plays important part in
the modern theories and applications. Optics of metamaterials,
transformation optics, and nonlinear optics are the basis for
constructing smart devices for effective light control. Together
with the materials the electromagnetic fields become also more
intricate. For example, accelerated Airy beams, nonparaxial Bessel
beams, and knotted fields propose the novel interesting physics
behind them \cite{Siviloglou,Chen,Novitsky,Sukhov,Irvine}. With more
complicated three-dimensional electromagnetic beams, the
generalizations\cite{Milione} of the degree of polarization and
coherence matrix may be appreciated.

The theory of polarization optics was developed in the aforetime
centuries: Poincare's sphere,\cite{Poincare} Stokes parameters,
\cite{Stokes} Wolf's coherence matrix,\cite{Wolf} $2 \times 2$ Jones
matrix, \cite{Jones} $4 \times 4$ Mueller matrix, \cite{Mueller}
etc. The coherence matrix serves for the description of a partially
polarized beam, when it propagates in a certain direction. In Refs.
\cite{Fedorov65,Fedorov} the coherence matrix was generalized to the
so called \emph{light beam tensor}, which keeps invariant with
respect to the rotations in the three-dimensional space. The
coherence matrix is the special representation of this tensorial
quantity.

In the present paper we investigate the light beam tensor for the
three-dimensional electromagnetic beams. In the previous studies,
the three-dimensional coherence matrix, 9 Stokes parameters, and
three-dimensional degree of polarization were
introduced.\cite{Setala2002,Setala2009} However, the proposed degree
of polarization is just the mathematical generalization of the
two-dimensional coherence matrix. Physically justified 3D degree of
polarization \cite{Ellis,EllisOC05,EllisOL04,Brosseau,Refregier} as
the ratio of the intensity of the fully polarized field to the total
intensity turns out to be different quantity.

In Section II of the paper we discuss the definition of the
unpolarized field and find out that the 3D coherence matrix is not
able to describe the general beam structure. In Section III we
derive that the mathematically generalized degree of polarization
\cite{Setala2002,Setala2009} coincides with physically defined one
\cite{Ellis} when the beam consists of completely polarized and
completely unpolarized components \cite{Aunon}. In this case we
study the light beam tensor in details. This beam tensor involves
only 6 independent parameters, therefore, requires 6 Stokes
parameters. In Section IV, the choice of the 6 Stokes parameters is
discussed. The problem of the reconstruction of the
three-dimensional light beam's tensor using the in-plane
measurements is considered in Section V. Section VI concludes the
paper.

\section{Unpolarized light}

When one characterizes 2D partially polarized light, it is
intuitively clear that the polarized light is the coherent
superposition of partial waves, while unpolarized light is
non-coherent superposition of partial waves which polarizations are
equiprobable. For the 3D light one naturally keeps the definition of
polarized wave \cite{EllisOL04}. Unpolarized light is not well
defined and treated as something complicated. In this Section we
justify the definition of the 3D unpolarized light as superposition
of equiprobably polarized non-coherent waves in the
three-dimensional space.

\begin{figure}[t!]
\centering \epsfig{file=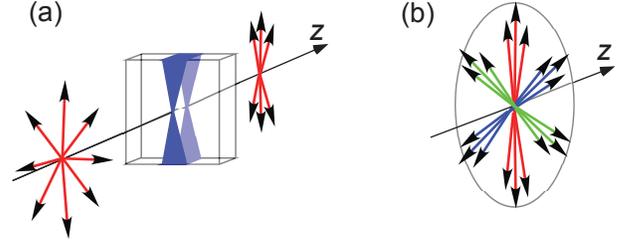, width=0.45\textwidth} \caption{
(a) Incident natural light is polarized within the angle sector in
the beam plane. (b) 2D beam generated as the superposition of beams
unpolarized in three angle sectors. } \label{Fig1}
\end{figure}

Let us start with the 2D unpolarized light. Is it really clear that
the partially polarized beam consists of completely polarized beam
and completely unpolarized beam and, therefore, can be described by
the 2D coherence matrix? Consider a device (incomplete polarizer)
that can transmit only the waves which polarizations belong to an
angle sector as indicated in Fig. \ref{Fig1}(a). Then the incident
naturally polarized beam becomes unpolarized light which
polarizations are equiprobable in this angle sector. When several
such beams are mixed (Fig. \ref{Fig1}(b)), the 2D state of
polarization cannot be fully described by the regular coherence
matrix. Should we say that the coherence matrix of the 2D field is
limited? We think it is just needed to consider usual definition of
the unpolarized light as superposition of equiprobable
polarizations.

\begin{figure}[t!]
\centering \epsfig{file=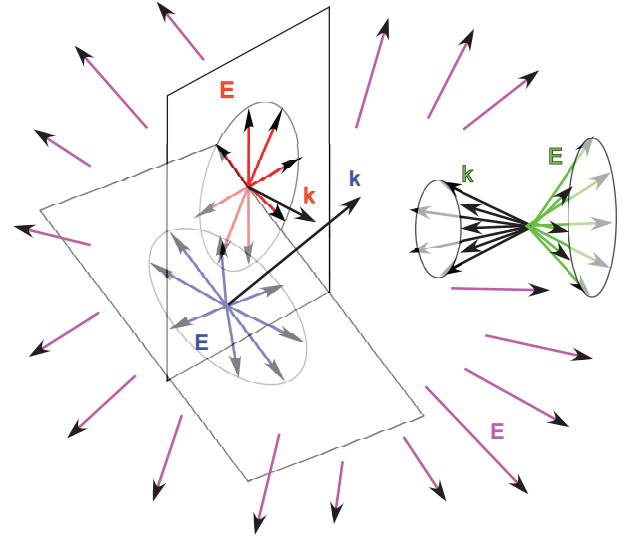, width=0.45\textwidth} \caption{ 3D
partly polarized field as superposition of 3D unpolarized beam
(electric fields are equiprobably directed), two in-plane
unpolarized beams, and on-cone unpolarized beam. Black arrows
indicate wavevectors ${\bf k}$, colored arrows show electric fields
$\ee$.  } \label{Fig2}
\end{figure}

The similar story is usually narrated for 3D electromagnetic fields.
It is accepted that a 3D partially polarized field can be presented
as superposition of completely polarized light, completely
unpolarized light (mix of equiprobably polarized waves), and
something else which is related to unpolarized light (2D unpolarized
light according to Ref. \cite{EllisOC05}). In general, electric
field of the 3D unpolarized light can be treated as the sum of
electric fields of different unpolarized electromagnetic beams, such
as in-plane completely unpolarized, completely unpolarized waves
with wavevectors lying on the cone (Bessel beams), etc. Some
examples are demonstrated in Fig. \ref{Fig2}. It is evident that if
the 3D unpolarized field is formed by many such partial beams, we
need much more than 9 parameters which are introduced to describe
the $3\times 3$ coherence matrix. Thus we fundamentally cannot
retrieve the realistic structure of the 3D field, if we know nothing
of the field.

Let us inspect the consequences of the above reasoning applied for
the coherence matrices (light beam tensors). Superposition of
non-coherent elementary plane waves $\ee^{(s)}$ with the same
direction of propagation ${\bf k}^{(s)}= k {\bf n}$, where $s$
enumerates the elementary waves, ${\bf k}^{(s)}$ is the wavevector,
$|{\bf n}| = 1$, is described by the light beam tensor
\cite{Fedorov65,Fedorov}
\begin{equation}
\Phi_2 = \sum_s \ee^{(s)} \otimes \ee^{(s)\ast}, \qquad \Phi_2 {\bf
n} = {\bf n} \Phi_2 = 0, \label{2D_bt}
\end{equation}
where ${\bf a} \otimes {\bf b}$ is the dyad (tensor product of the
vectors ${\bf a}$ and ${\bf b}$). In the index form, $({\bf a}
\otimes {\bf b})_{ij} = a_i b_j$, $i,j=1,2,3$. Quantity $\Phi_2$
defined by Eq. (\ref{2D_bt}) is indeed tensor, since it is composed
of elementary tensors, dyads.

Three-dimensional light represents the superposition of elementary
waves, which can possess not only random phases and polarizations,
but also \emph{directions of propagation}, i.e. ${\bf k}^{(s)}=
k^{(s)} {\bf n}^{(s)}$. Thus, the light beam tensor equals
\begin{equation}
\Phi = \Phi_3 = \sum_s \ee^{(s)} \otimes \ee^{(s)\ast}.
\label{3D_bt}
\end{equation}
In contrast to Eq. (\ref{2D_bt}), the additional limitations on the
beam tensor $\Phi {\bf n} = {\bf n} \Phi = 0$ are not valid anymore.

When the beam consists of coherent elementary waves, it is fully
polarized and the beam's tensor $\Phi = \ee \otimes \ee^\ast$. In
the opposite situation of 3D fully unpolarized light with
equiprobable polarizations the beam tensor does not have a preferred
direction and, therefore, $\Phi = A \textbf{1}$, where ${\bf 1}$ is
the identity tensor in the three-dimensional space and $A$ is a
coefficient.

As any self-conjugated tensor ($\Phi^\dag = \Phi$, here $\dag$
denotes Hermitian conjugate), three-dimensional beam's tensor can be
presented as a spectral expansion of the form
\begin{equation}
\Phi = \lambda_1 {\bf u}_1 \otimes {\bf u}_1^\ast + \lambda_2 {\bf
u}_2 \otimes {\bf u}_2^\ast + \lambda_3 {\bf u}_3 \otimes {\bf
u}_3^\ast, \label{spectralPhi}
\end{equation}
where $\lambda_i = \lambda_i^\ast$ and ${\bf u}_i$ ($i = 1,2,3$) are
the eigenvalues and eigenvectors of $\Phi$, respectively. The
eigenvectors are orthogonal and normalized as ${\bf u}_i {\bf
u}^\ast_j = \delta_{ij}$, where $\delta_{ij}$ is Kronecker's delta.
Spectral decomposition for the identity tensor reduces to the
completeness condition ${\bf 1} = {\bf u}_1 \otimes {\bf u}_1^\ast +
{\bf u}_2 \otimes {\bf u}_2^\ast + {\bf u}_3 \otimes {\bf
u}_3^\ast$. Tensor $\Phi$ can be described by 9 independent
parameters.

On the other hand, the general definition (\ref{3D_bt}) can be
written in the form different from Eq. (\ref{spectralPhi}), if we
know something of the electromagnetic beam. For example, let the 3D
field includes the superposition of 2D completely unpolarized beams
with directions defined by the unit vectors ${\bf n}_\alpha$
($\alpha = 1,\ldots, M$). Then the beam tensor reads
\begin{equation}
\Phi = {\bf E}_p \otimes {\bf E}_p^\ast + A {\bf 1} +
\sum_{\alpha=1}^M B_\alpha I_{\alpha}, \label{BT_example1}
\end{equation}
where $I_{\alpha} = {\bf 1} - {\bf n}_\alpha \otimes {\bf n}_\alpha$
is the projector onto the plane with normal vector ${\bf n}_\alpha$.
If the directions of the normal vectors are known, the tensor Eq.
(\ref{BT_example1}) depends on 5 parameters of polarized field
(${\bf E}_p \otimes {\bf E}_p^\ast$), 1 parameter of 3D fully
unpolarized beam ($A {\bf 1}$), and $M$ parameters of 2D unpolarized
beams ($\sum_{\alpha=1}^M B_\alpha I_{\alpha}$). If $6+M > 9$, we
cannot reconstruct the structure of the beam using 9 parameters of
the general 3D coherence matrix Eq. (\ref{spectralPhi}).
Nevertheless, the beam parameters can be found, if we make more
measurements than 9. But this can be done only if we know the form
of the coherence matrix, e.g. Eq. (\ref{BT_example1}).

If the directions of propagation of the 2D unpolarized beams are
unknown, it is necessary to introduce two additional parameters for
each real unit vector ${\bf n}_\alpha$, i.e. the number of unknown
parameters is equal to $6+3M$. When $M=1$, the coherence matrix
indeed can be presented as the sum of coherent part, completely 3D
unpolarized part, and completely 2D unpolarized part, respectively:
\begin{equation}
\Phi = {\bf E}_p \otimes {\bf E}_p^\ast + A {\bf 1} + B ({\bf 1} -
{\bf n} \otimes {\bf n}). \label{BT_example2}
\end{equation}
If the directions of propagation of 2D unpolarized beams are known,
it is feasible to specify 3D field using the coherence matrix Eq.
(\ref{BT_example1}) for $M \le 3$.

Concluding this section, it is not possible to determine the actual
structure of either 3D or 2D field, if we know nothing about the
field itself, because the coherence matrix does not carry sufficient
information. In 2D case, it is accepted that the fully unpolarized
light possesses equiprobable in-plane polarizations. We are
convinced that the similar definition of the completely unpolarized
light should be used for the 3D light. If the beam is arbitrary, it
is not sufficient to set even 9 components of the coherence matrix
(\ref{spectralPhi}) to \emph{unambiguously} determine the structure
of the beam.

\section{Three-dimensional light}

\subsection{Form of light beam's tensor}

According to the results of the previous Section we consider the 3D
field as composed of completely 3D polarized light and completely 3D
unpolarized light. This means that there are no specific directions
and two forms of light beam tensor
\begin{equation}
\Phi = (\lambda_1 - \lambda_3) {\bf u}_1 \otimes {\bf u}_1^\ast +
(\lambda_2 - \lambda_3) {\bf u}_2 \otimes {\bf u}_2^\ast + \lambda_3
{\bf 1} \label{PhiLambda}
\end{equation}
and
\begin{equation}
\Phi = (\lambda_1 - \lambda_2) {\bf u}_1 \otimes {\bf u}_1^\ast +
(\lambda_3 - \lambda_2) {\bf u}_3 \otimes {\bf u}_3^\ast + \lambda_2
{\bf 1} \label{PhiLambda1}
\end{equation}
should be equivalent. Thus we conclude that $\lambda_2 = \lambda_3$
and the beam tensor of the three-dimensional light equals
\begin{equation}
\Phi = (\lambda_1 - \lambda_2) {\bf u}_1 \otimes {\bf u}_1^\ast +
\lambda_2 {\bf 1}, \label{PhiFinal}
\end{equation}
where $\Phi_p = (\lambda_1 - \lambda_2) {\bf u}_1 \otimes {\bf
u}_1^\ast$ and $\Phi_u = \lambda_2 {\bf 1}$ describe completely
polarized and unpolarized light, respectively.

3D field is completely polarized, when eigenvalue $\lambda_2 = 0$,
and completely unpolarized, when $\lambda_1 = \lambda_2$. For all
other values $0 \le \lambda_2 \le \lambda_1$ one gets partially
polarized light. Eq. (\ref{PhiFinal}) has clear physical meaning,
because it includes intuitively defined polarized and unpolarized
beam's tensors. The form of the three-dimensional beam tensor
(\ref{PhiFinal}) can be formulated using another argumentation. In
the three-dimensional space, there is the single distinguished
direction of the polarized electric field ${\bf u}_1$. Therefore, we
can construct only the tensor of the form $\alpha {\bf u}_1 \otimes
{\bf u}_1^\ast + \beta {\bf 1}$.

Two-dimensional light beam tensor is characterized by two
distinguished directions, ${\bf u}_1$ and ${\bf n}$, and can be
presented in the similar form as \cite{Fedorov65,Fedorov}
\begin{equation}
\Phi_2 = (\lambda_1 - \lambda_2) {\bf u}_1 \otimes {\bf u}_1^\ast +
\lambda_2 I, \label{PhiFinal2D}
\end{equation}
where $I = {\bf 1} - {\bf n} \otimes {\bf n}$ is the projection
operator onto the plane with normal vector ${\bf n}$ and ${\bf u}_1
= I {\bf u}_1$ is the vector in the plane orthogonal to ${\bf n}$.
Eq. (\ref{PhiFinal}) differs from Eq. (\ref{PhiFinal2D}) with the
three-dimensional vector ${\bf u}_1$ and three-dimensional identity
tensor.

\subsection{Degree of polarization}

Eq. (\ref{PhiFinal}) provides intuitive definition of the
three-dimensional degree of polarization in terms of the eigenvalues
of the beam tensor. The trace of the coherence matrix ${\rm
Tr}(\Phi)$ is proportional to the intensity of light. Intensity of
the polarized and unpolarized beams are ${\rm Tr}(\Phi_p) =
(\lambda_1 - \lambda_2)$ and ${\rm Tr}(\Phi_u) = 3 \lambda_2$,
respectively. Degree of polarization for the three-dimensional light
is equal to
\begin{equation}
P_3 = \frac{{\rm Tr}(\Phi_p)}{{\rm Tr}(\Phi_p) + {\rm Tr}(\Phi_u)} =
\frac{ \lambda_1 - \lambda_2}{\lambda_1 + 2\lambda_2},
\label{DoPlambda}
\end{equation}

The two eigenvalues of the coherence matrix can be found using the
two invariants of $\Phi$. Usually the trace of the matrix ${\rm
Tr}(\Phi) \equiv (\Phi)_t$ and the trace of the squared matrix ${\rm
Tr}(\Phi^2) \equiv (\Phi^2)_t$ are used. From $\Phi = \lambda_1 {\bf
u}_1 \otimes {\bf u}_1^\ast + \lambda_2 ({\bf u}_2 \otimes {\bf
u}_2^\ast + {\bf u}_3 \otimes {\bf u}_3^\ast)$ one easily derives
\begin{equation}
(\Phi)_t = \lambda_1 + 2 \lambda_2, \qquad (\Phi^2)_t = \lambda^2_1
+ 2 \lambda^2_2. \label{invariants}
\end{equation}

For $\lambda_2 \le \lambda_1$ we obtain
\begin{eqnarray}
\lambda_1 = \frac{1}{3} \left( (\Phi)_t + \sqrt{6 (\Phi^2)_t - 2
(\Phi)_t^2} \right), \nonumber \\
\lambda_2 = \frac{1}{6} \left( 2 (\Phi)_t - \sqrt{6 (\Phi^2)_t - 2
(\Phi)_t^2} \right). \label{labmda12}
\end{eqnarray}
Degree of polarization takes the form
\begin{equation}
P_3 = \sqrt{\frac{3}{2} \frac{(\Phi^2)_t}{(\Phi)_t^2} -
\frac{1}{2}}. \label{DoPinvar}
\end{equation}
For completely polarized light $P_3 = 1$ and $(\Phi^2)_t =
(\Phi)_t^2$. Completely unpolarized light is characterized by $P_3 =
0$ and $(\Phi^2)_t = (\Phi)_t^2/3$. Thus, the degree of polarization
is in the interval $0 \le P_3 \le 1$.

It should be noted that the generalized degree of polarization
obtained in Ref. \cite{Setala2002} coincides with Eq.
(\ref{DoPinvar}). This means that the generalization of the 2D
degree of polarization inherits the property of partially polarized
beam to be split into completely polarized and unpolarized parts. In
other words, the degree of polarization in Ref. \cite{Setala2002}
corresponds to the restricted coherence matrix Eq. (\ref{PhiFinal}).
When the coherence matrix is the general $3\times 3$ Hermitian
matrix, the degree of polarization is expressed as $(\lambda_1 -
\lambda_2)/(\lambda_1+\lambda_2+\lambda_3)$ (see Ref. \cite{Ellis}).
In this case, the three eigenvalues can be found using three
invariants of the coherence matrix ${\rm Tr}(\Phi)$, ${\rm
Tr}(\Phi^2)$, and ${\rm det}(\Phi)$, and closed-form expression for
$P_3$ is expected to be more complicated. For the derived beam
tensor (\ref{PhiFinal}) the generalized and physically justified
degrees of polarization are agreed as it has been pointed out in
Ref.\cite{Aunon}

\subsection{From 3D to 2D degree of polarization}

Transition from the three-dimensional light to the two-dimensional
one can be performed by excluding one eigenvector (assuming, e.g.,
${\bf u}_3 = 0$). In the previously derived formulae we have
considered vector ${\bf u}_3$ normalized by unity, what should be
violated in the 2D case. So, we will explicitly write the vector
${\bf u}_3$ in equations. Then ${\rm Tr}(\Phi_p) = (\lambda_1 -
\lambda_2)$ and ${\rm Tr}(\Phi_u) = \lambda_2 (2 + |{\bf u}_3|^2)$,
and
\begin{equation}
P_3 = \frac{ \lambda_1 - \lambda_2}{\lambda_1 + \lambda_2 (1 + |{\bf
u}_3|^2)}. \label{DoPlambda1}
\end{equation}

In terms of the invariants of the coherence matrix, the degree of
polarization of the beam takes the form
\begin{equation}
P_3 = \alpha_0 + \sqrt{\alpha_1 \frac{(\Phi^2)_t}{(\Phi)_t^2} -
\alpha_2}, \label{DoPinvar1}
\end{equation}
where
\begin{eqnarray}
\alpha_0 &=& \frac{|{\bf u}_3|^2 (|{\bf u}_3|^2 -1)}{2 (1 + |{\bf
u}_3|^2 + |{\bf u}_3|^4)}, \qquad
\alpha_1 = \frac{(2 + |{\bf u}_3|^2)^2}{2 (1 + |{\bf
u}_3|^2 + |{\bf u}_3|^4)}, \nonumber \\
\alpha_2 &=& \frac{(2 + |{\bf u}_3|^2)^2 (1 + |{\bf u}_3|^4)}{4 (1 +
|{\bf u}_3|^2 + |{\bf u}_3|^4)^2}. \label{coefficAlpha}
\end{eqnarray}

Eqs. (\ref{DoPlambda1}) and (\ref{DoPinvar1}) are valid both for
three-dimensional and two-dimensional fields. For 3D and 2D fields
one needs to apply $|{\bf u}_3|^2 = 1$ and $|{\bf u}_3|^2 = 0$,
respectively.

\subsection{From 3D to 2D light beam's tensor}

Three-dimensional beam's tensor (\ref{3D_bt}) is the sum of the
dyads $\ee^{(s)} \otimes \ee^{(s)\ast}$. When we want to study the
light on a plane, we need to consider projected fields $I_{\bf m}
\ee^{(s)}$, where $I_{\bf m} = {\bf 1} - {\bf m} \otimes {\bf m}$ is
the projector onto the plane with normal vector ${\bf m}$ ($|{\bf
m}| =1$). The beam tensor composed of such projected electric fields
equals
\begin{equation}
\Phi_2({\bf m}) = \sum_s I_{\bf m} \ee^{(s)} \otimes I_{\bf m}
\ee^{(s)\ast} = I_{\bf m} \Phi I_{\bf m}. \label{Phi2Im}
\end{equation}
When the elementary waves of the beam propagate in the same
direction ${\bf n}$, we obtain the ordinary two-dimensional
coherence matrix
\begin{equation}
\Phi_2 = I_{\bf n} \Phi I_{\bf n} = (\lambda_1 - \lambda_2) {\bf
u}_1 \otimes {\bf u}_1^\ast + \lambda_2 I_{\bf n}.
\end{equation}
Here $I_{\bf n} {\bf u}_1 = {\bf u}_1$ is the electric field vector
in the plane of constant phase. It should be noted that in general
we can write the beam tensor projection on \emph{any} plane
according to Eq. (\ref{Phi2Im}).

\section{Stokes parameters}

If $\Phi$ was the general matrix (\ref{spectralPhi}), it would
contain 9 independent parameters, which could be written as Stokes
parameters for the three-dimensional fields. However, the coherence
matrix has reduced form (\ref{PhiFinal}), which decreases the number
of independent parameters.

Let us calculate the number of independent parameters of beam's
tensor (\ref{PhiFinal}). Normalized complex vector ${\bf u}_1$ can
be expressed in terms of 4 real quantities, $a$, $b$, $\varphi_1$,
and $\varphi_2$, as
\begin{equation}
{\bf u}_1 = a {\bf e}_x + b {\rm e}^{i \varphi_1} {\bf e}_y +
\sqrt{1 - a^2 - b^2} {\rm e}^{i \varphi_2} {\bf e}_z. \label{u1}
\end{equation}
(Coefficient in front of ${\bf e}_x$ can be regarded as real,
because ${\bf u}_1$ enters beam's tensor as ${\bf u}_1 \otimes {\bf
u}_1^\ast$.) Adding two more real eigenvalues $\lambda_1$ and
$\lambda_2$, we claim 6 independent parameters for the light beam
tensor.

Definition of the three-dimensional beam tensor in the form
(\ref{PhiFinal}) does not take into account the transversality
condition $\nabla \ee = 0$. Completely polarized electric field
$\ee$ can be found from the definition  $\Phi_p = \ee \otimes
\ee^\ast = (\lambda_1 - \lambda_2) {\bf u}_1 \otimes {\bf u}_1^\ast
$. Then the electric field equals $\ee = \exp(i \psi)
\sqrt{\lambda_1 - \lambda_2} {\bf u}_1$, while the transversality
condition reads
\begin{equation}
\nabla( {\rm e}^{i \psi} \sqrt{\lambda_1 - \lambda_2} {\bf u}_1) =
0. \label{transverse}
\end{equation}
In general, the phase $\psi({\bf r})$ distribution cannot be
supposed and Eq. (\ref{transverse}) is the differential equation for
the phase $\psi$:
\begin{equation}
i \sqrt{\lambda_1 - \lambda_2} ({\bf u}_1 \nabla\psi) + \nabla
(\sqrt{\lambda_1 - \lambda_2} {\bf u}_1) = 0.
\end{equation}
The number of independent parameters for $\Phi$ is still 6.

When we know the phase, e.g., $\psi = \beta z$ ($\beta$ is the
propagation constant of the beam), the transversality condition
\begin{equation}
- \beta \sqrt{\lambda_1 - \lambda_2} ({\bf e}_z {\bf u}_1) + \nabla
(\sqrt{\lambda_1 - \lambda_2} {\bf u}_1) = 0 \label{transverse1}
\end{equation}
becomes the pair of restrictions on $\lambda_{1,2}$ and ${\bf u}_1$
of the form ${\rm Re}(\nabla \ee) = 0$ and ${\rm Im}(\nabla \ee) =
0$, so that the beam is fully described by the 4 independent
parameters (Stokes parameters). When the beam consists of the plane
waves propagating in the direction of vector ${\bf n}$, $\psi({\bf
r}) = k ({\bf n} {\bf r})$ and $\sqrt{\lambda_1 - \lambda_2} {\bf
u}_1$ is constant. Eq. (\ref{transverse1}) reads ${\bf n} {\bf u}_1
= 0$ or ${\bf n} {\bf E} = 0$. This is equivalent to the conditions
$\Phi_2 {\bf n} = {\bf n} \Phi_2 = 0$ on the coherence matrix used
in the definition (\ref{2D_bt}).

Thus, if there are no preferred directions, the transversality
condition just exhibits the differential equation for the phase
$\psi$ and does not decrease the number of the independent
parameters $a$, $b$, $\varphi_1$, $\varphi_2$, $\lambda_1$, and
$\lambda_2$. If the phase $\psi$ is somehow defined, there are two
additional equations for the parameters, and we can use only $a$,
$\varphi_1$, $\lambda_1$, and $\lambda_2$. So, we should have 6
Stokes parameters for truly three-dimensional fields and 4 Stokes
parameters for 2D fields, when we can introduce the preferred
direction (say, the direction of the beam propagation).

The Stokes parameters can be introduced as it was done in Ref.
\cite{Setala2002} but then we need to choose only 6 parameters of 9,
which are independent. The rest 3 parameters can be expressed using
the independent 6 parameters. One of such links between the Stokes
parameters is shown below:
\begin{equation}
\frac{\Lambda_1^2 + \Lambda_2^2}{\Lambda_3^2} \left[
\frac{\Lambda_4^2 + \Lambda_5^2}{\Lambda_6^2 + \Lambda_7^2} - 1
\right]^2 = 4 \frac{\Lambda_4^2 + \Lambda_5^2}{\Lambda_6^2 +
\Lambda_7^2},
\end{equation}
where $\Lambda_j$ ($j = 0,\ldots,8$) are the Stokes parameters
introduced in Ref. \cite{Setala2002} for the three-dimensional
fields. Two more links can be derived.

However, since most of the Stokes parameters $\Lambda_j$ have no
physical sense and can be found from the coherence matrix, we
propose another set of the Stokes parameters. For example, it is
more convenient to use 4 conventional Stokes parameters in some
plane (e.g., in ($x$, $y$) plane) and two more parameters in another
plane (e.g., in ($x$, $z$) plane):
\begin{eqnarray}
S_0 &=& \Phi_{xx} + \Phi_{yy}, \qquad S_1 = \Phi_{xx} - \Phi_{yy},
 \nonumber \\
S_2 &=& \Phi_{xy} + \Phi_{yx}, \qquad S_3 = i(\Phi_{yx} -
\Phi_{xy}),
\nonumber \\
S_4 &=& \Phi_{xz} + \Phi_{zx}, \qquad S_5 = i (\Phi_{zx} -
\Phi_{xz}).
\end{eqnarray}
Then $S_0$ is proportional to the field intensity in the plane of
detector and $S_3$ is proportional to the spin angular momentum in
the $z$-direction.

From the point of view of physics the field intensities and spin
angular momenta are beneficial as independent parameters. Therefore,
we also propose the \emph{physical Stokes parameters} for the
three-dimensional beams:
\begin{eqnarray}
S'_0 &=& S_0 = \Phi_{xx} + \Phi_{yy}, \qquad S'_1 = \Phi_{xx} +
\Phi_{zz},
 \nonumber \\
S'_2 &=& \Phi_{yy} + \Phi_{zz}, \qquad S'_3 = S_3 = i(\Phi_{yx} -
\Phi_{xy}),
\nonumber \\
S'_4 &=& i (\Phi_{zx} - \Phi_{xz}), \qquad S'_5 = i(\Phi_{zy} -
\Phi_{yz}).
\end{eqnarray}
Parameters $S'_0$, $S'_1$, and $S'_2$ stand for the intensities in
the planes ($x$, $y$), ($x$, $z$), and ($y$, $z$), respectively,
while $S'_3$, $S'_4$, and $S'_5$ describe the spin angular momenta
in directions $z$, $y$, and $x$. $S'_0$ and $S'_3$ coincide with
analogous quantities of the usual set of Stokes parameters $S_j$
($j=0,1,2,3$). With the eigenvalues $\lambda_{1,2}$ and 4 parameters
of the vector ${\bf u}_1$ (see Eq. (\ref{u1})), the physical Stokes
parameters take the form
\begin{eqnarray}
S'_0 &=& \lambda_p(a^2 + b^2) + \lambda_2, \quad S'_1 =
\lambda_p(a^2 + c^2) + \lambda_2,
 \nonumber \\
S'_2 &=& \lambda_p(b^2 + c^2) + \lambda_2, \quad S'_3 = - 2
\lambda_p a b \sin(\varphi_1),
\nonumber \\
S'_4 &=& - 2 \lambda_p a c \sin(\varphi_2), \quad S'_5 = - 2
\lambda_p b c \sin(\varphi_2 - \varphi_1),
\end{eqnarray}
where $\lambda_p = \lambda_1 - \lambda_2$ and $c = \sqrt{1- a^2 -
b^2}$. The two-dimensional Stokes parameters require $c = 0$ or $a^2
+ b^2 = 1$. In this case we have only 4 independent parameters
$\lambda_{1,2}$, $a$, and $\varphi_1$, and the physical Stokes
parameters
\begin{eqnarray}
S'_0 &=& \lambda_p + \lambda_2, \quad S'_1 = \lambda_p a^2 +
\lambda_2,
 \nonumber \\
S'_2 &=& \lambda_p (1 -a^2) + \lambda_2, \quad S'_3 = - 2 \lambda_p
a b \sin(\varphi_1),
\nonumber \\
S'_4 &=& 0, \quad S'_5 = 0
\end{eqnarray}
are reduced to the four quantities, which can be connected with the
ordinary Stokes parameters as
\begin{eqnarray}
S_0 &=& S'_0, \qquad S_1 = S'_1 - S'_2,
 \nonumber \\
S_2 &=& \sqrt{4 (S'_0 - S'_1) (S'_0 - S'_2) - S_3^{\prime 2}},
\qquad S_3 = S'_3.
\end{eqnarray}

\section{Reconstruction of 3D coherence matrix}

Measurement of nine components of the coherence matrix for 3D fields
were discussed in Ref. \cite{EllisPRL05}. Here we deal with the
measurement of the components of the coherence matrix Eq.
(\ref{PhiFinal}) using the measurement of the fields in the detector
plane. Let us denote the normal vector to the detector plane as
${\bf m}$ and reveal how the position of this plane influences the
beam's tensor and degree of polarization.

\subsection{Projected 3D beam's tensor}

We will determine characteristics of the 2D beam tensor as
projection of the 3D beam tensor. The absolute value of the
three-dimensional vector ${\bf u}_1$ projected on a plane is less
than unity and we lose the information about vector component
orthogonal to the detector ${\bf m}{\bf u}_1$. Then projected beam's
tensor
\begin{equation}
\Psi_2 = I_{\bf m} \Phi_3 I_{\bf m} = (\lambda_1 - \lambda_2)
(I_{\bf m} {\bf u}_1) \otimes (I_{\bf m} {\bf u}_1^\ast) + \lambda_2
I_{\bf m}
\end{equation}
can be rewritten in the form
\begin{equation}
\Psi_2 = (\lambda_1 - \lambda_2) |I_{\bf m} {\bf u}_1|^2 {\bf v}
\otimes {\bf v}^\ast + \lambda_2 I_{\bf m},
\end{equation}
where ${\bf v} = (I_{\bf m} {\bf u}_1)/|I_{\bf m} {\bf u}_1|$ is
situated in the plane with the normal vector ${\bf m}$, $|{\bf
v}|=1$, and $I_{\bf m}$ is the projection operator. The
detector-measured intensities of the completely polarized and
unpolarized parts of the beam are $(\lambda_1 - \lambda_2) |I_{\bf
m} {\bf u}_1|^2$ and $2 \lambda_2$, respectively. Degree of
polarization is defined as the part of the intensity of completely
polarized beam divided by the total intensity:
\begin{equation}
P_2 = \frac{(\lambda_1 - \lambda_2) |I_{\bf m} {\bf
u}_1|^2}{(\lambda_1 - \lambda_2) |I_{\bf m} {\bf u}_1|^2 + 2
\lambda_2}.
\end{equation}
$\lambda_{1,2}$ are defined by Eq. (\ref{labmda12}) via invariants
of the 3D coherence matrix. In terms of invariants of $\Psi_2$ we
will get to the usual formula for the degree of polarization at the
plane: $P_2 = \sqrt{2 (\Psi_2^2)_t/(\Psi_2)_t^2 - 1}$.

\subsection{Retrieval procedure}

\begin{figure}[t!]
\centering \epsfig{file=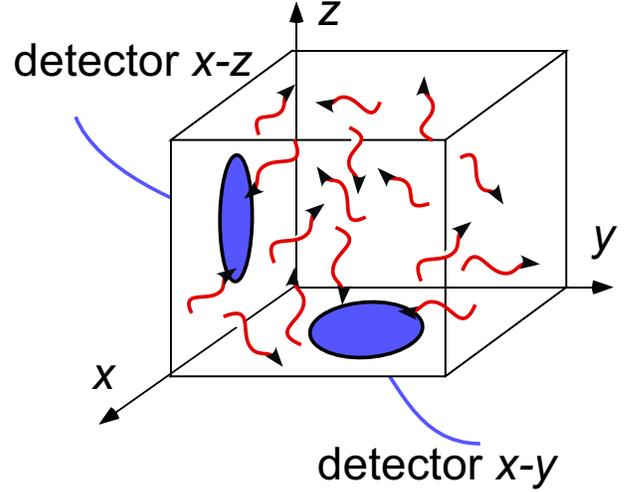, width=0.45\textwidth} \caption{ 3D
electromagnetic field in a cavity. Detectors are placed on the ($x$,
$y$) and ($x$, $z$) planes. } \label{Fig3}
\end{figure}

We aim to retrieve the 3D beam's tensor using detectors
(measurements in plane, see Fig. \ref{Fig3}). At first one can
measure the ordinary 2D coherence matrix in some plane (name it
($x$, $y$) plane)
\begin{equation}
\Psi_2^{(z)} = (\lambda_1 - \lambda_2) (I_z {\bf u}_1) \otimes (I_z
{\bf u}_1^\ast) + \lambda_2 I_z.
\end{equation}

As a result, we determine
\begin{eqnarray}
%&& (\lambda_1 - \lambda_2) |I_z {\bf u}_1|^2 = \sqrt{2
%(\Psi_2^{(z)2})_t - (\Psi_2^{(z)})_t^2}, \nonumber \\
%
\lambda_2 &=& \frac{1}{2} \left( (\Psi_2^{(z)})_t - \sqrt{2
(\Psi_2^{(z)2})_t - (\Psi_2^{(z)})_t^2} \right), \nonumber \\
I_z {\bf u}_1 &=& \sqrt{\frac{(\Psi^{(z)}_2)_{xx} -
\lambda_2}{\lambda_1 - \lambda_2}} {\bf e}_x \nonumber \\
&+& \frac{(\Psi^{(z)}_2)_{yx}}{\sqrt{(\lambda_1 - \lambda_2)
((\Psi^{(z)}_2)_{xx} - \lambda_2)}} {\bf e}_y
\end{eqnarray}
by means of the known matrix $\Psi_2^{(z)}$. From these equations we
unambiguously find $\lambda_{2}$ and direction of the vector $I_z
{\bf u}_1/|I_z {\bf u}_1|$.

Complete retrieval of the 3D coherence matrix requires knowing the
vector in 3D space, i.e. measurements out of the plane ($x$, $y$).
If we put a detector in the plane ($x$, $z$), using the coherence
matrix
\begin{equation}
\Psi_2^{(y)} = (\lambda_1 - \lambda_2) (I_y {\bf u}_1) \otimes (I_y
{\bf u}_1^\ast) + \lambda_2 I_y
\end{equation}
we can write another projection of the vector ${\bf u}_1$
\[
I_y {\bf u}_1 = \sqrt{\frac{(\Psi^{(y)}_2)_{xx} -
\lambda_2}{\lambda_1 - \lambda_2}} {\bf e}_x +
\frac{(\Psi^{(y)}_2)_{zx}}{\sqrt{(\lambda_1 - \lambda_2)
((\Psi^{(y)}_2)_{xx} - \lambda_2)}} {\bf e}_z.
\]
Then the vector under search is
\begin{eqnarray}
{\bf u}_1 = I_z {\bf u}_1 + {\bf e}_z ({\bf e}_z I_y {\bf u}_1)
\end{eqnarray}
or in the explicit form
\begin{eqnarray}
{\bf u}_1 = \sqrt{\frac{(\Psi^{(z)}_2)_{xx} - \lambda_2}{\lambda_1 -
\lambda_2}} {\bf e}_x + \frac{(\Psi^{(z)}_2)_{yx}}{\sqrt{(\lambda_1
- \lambda_2) ((\Psi^{(z)}_2)_{xx} - \lambda_2)}} {\bf e}_y \nonumber \\
+ \frac{(\Psi^{(y)}_2)_{zx}}{\sqrt{(\lambda_1 - \lambda_2)
((\Psi^{(y)}_2)_{xx} - \lambda_2)}} {\bf e}_z.
\end{eqnarray}
$\lambda_2$ has been found from the in-plane measurement, while
$\lambda_1$ (and the final form of ${\bf u}_1$) follows from the
normalization condition $|{\bf u}_1| = 1$ as
\begin{eqnarray}
\lambda_1 = \lambda_2 + |(\Psi^{(z)}_2)_{xx} - \lambda_2| +
\frac{|(\Psi^{(z)}_2)_{yx}|^2}{|(\Psi^{(z)}_2)_{xx} - \lambda_2|} +
\frac{|(\Psi^{(y)}_2)_{zx}|^2}{ |(\Psi^{(y)}_2)_{xx} - \lambda_2|}.
\nonumber \\
\end{eqnarray}
Thus we reconstruct the three-dimensional beam tensor (coherence
matrix) described by Eq. (\ref{PhiFinal}). The degree of
polarization for the three-dimensional light follows from Eq.
(\ref{DoPinvar}).

\section{Conclusion}

We have derived the light beam tensor (coherence matrix) for
three-dimensional fields grounding on the expected symmetry
properties: beam's tensor should not change for rotations with
respect to the direction of the fully polarized light. Such a
restricted form of the coherence matrix is justified by the
definition of the fully polarized light as sum of equiprobably
polarized elementary waves. General form of coherence matrix deals
with the limited number of unpolarized beams and does not allow
determining the actual structure of the field in principle. That is
why it is not a great simplification to treat partially polarized
light as superposition of coherent field and 3D random field. All
the more so considered beam tensor (\ref{PhiFinal}) has clear
meaning and can be described by 6 independent parameters --- Stokes
parameters. For the three dimensional light it is natural to choose
6 physical parameters: 3 intensities in different planes and 3 spin
angular momenta. We call these values ``physical Stokes
parameters.'' Finally, we have developed the procedure of
reconstruction of the 3D light beam tensor using the measurements of
the 2D coherence matrices of projected fields.

In some situations it may be not really necessary to introduce the
3D coherence matrix at all. Indeed, if we define the plane of
detector as distinguished interface (like the plane of constant
phase in 2D case), then we can calibrate the degree of polarization
for the 3D beams with respect to this interface. Some different
beams may have the same degree of polarizations, though the beams
should be different. Only if it is crucial for the results, the full
reconstruction of the 3D coherence matrix and calculation of the 3D
degree of polarization is needed.

%\begin{acknowledgments}
%\end{acknowledgments}
\bigskip

%\appendix
%\section{ \label{appendix:a}}

\end{document}